\documentclass[aps,reprint]{revtex4-1}

\usepackage{epsfig,graphics,color}
\usepackage{graphicx}
\usepackage{dcolumn}
\usepackage{bm}

\begin{document}

\title{Point Charge Dynamics Near a Grounded Conducting Plane}

\author{Kevin L. Haglin}
\email[]{klhaglin@stcloudstate.edu}
\homepage[]{http://www.feynman.stcloudstate.edu/haglin/}
\affiliation{Department of Physics and Astronomy, Saint Cloud State University,
720 Fourth Avenue South, St. Cloud, MN 56301}

\date{\today}

\begin{abstract}
The classic image problem in electromagnetism involves a grounded 
infinite conducting plane and a point charge.  The force of attraction 
between the point charge and the plane is identified using an
equivalent-field picture of an image charge with opposite sign equidistant 
behind the plane resulting in a 1/{\it\/r\/}$^{\rm\/2}$ force of
attraction between the original charge and the plane.  If the 
point charge is released from rest it will reach the plane in a time $\tau$.
This time $\tau$ has not been calculated correctly 
up to now.  Clarification of the inconsistency is presented along
with a correct solution to the classic image problem.  Other 
electromagnetism problems are mentioned with attractive 
1/{\it\/r}$^{\it\,n}$ forces
(where {\it\/n\/} $\ge$ 1).  Such 
situations arise between a point charge and a line charge (1/{\it\/r\/}), 
between a line charge and a point dipole (1/{\it\/r\/}$^{\/2}$),
between
a point charge and a point dipole (1/{\it\/r\/}$^{\/3})$, and between 
a point dipole and a second point dipole (1/{\it\/r\/}$^{\/4}$).
\end{abstract}


\maketitle 

\section{Introduction}
\label{intro}

Point masses under the influence of Newtonian
gravity and point charges under the influence of Coulombic attraction
experience a 1/{\it\/r\/}$^{\rm\,2}$ force.
Particles in this
picture will approach one another according to Newtonian
(nonrelativistic) dynamics and if released from rest will reach the force
center in a time that is typically shown to be proportional to 
the initial separation to the three-halves power.  
This type of solution appears in
various places in textbooks as an application of gravitational
forces\cite{hrk02,thorntonmarion}, and in electromagnetism textbooks
surrounding the classic image problem\cite{griffiths,baymanhammermesh,palit}.
Unfortunately, the solution leading to this result is plagued
with superluminal ({\it\/i.e.\/} faster than light speed) transport and hence
cannot be formally correct.  

In addition to relativistic effects, action-at-a-distance type forces 
require propagation of information about the force.  News of changing 
particle positions must propagate to 
the other particle before any changes can be realized and the
news travels at a fixed (finite) speed, namely, the speed 
of light.  Care must be taken to properly account for these retardation 
effects.  The ultimate effect is to extend the time it takes for the particle
to reach the force center as compared with an assumption of instantaneous
updating. 

The electromagnetic category of problem is discussed
in this paper, but the same formalism can be applied to the gravitational
problem as well.  The solution discussed below for problems
of this type incorporates Einstein's special relativity and includes 
the effects of field retardation.
In Section~\ref{imageproblem} the
classic image problem is discussed and solved taking special 
relativity into account.  After all, when particle separation
approaches zero with an inverse squared force law, infinite kinetic
energies are possible.  This has all the earmarks of superluminal
dynamics.  Consistent results without any superluminal dynamics are presented 
in Section~(\ref{imageproblem}).
The Li\'enard-Wiechart potentials are employed in Sect.~(\ref{lienardwiechart})
to take into account the retardation effects.  Sect.~(\ref{lienardwiechart})
also includes a brief discussion to illustrate conditions 
under which the differences between the old (wrong) solution and the
new (correct) solution are important.  
Discussion of the correspondence to analagous gravitational problems
is included
in Section~(\ref{discuss}).  A conclusion
is included to summarize the main findings and to introduce other
problems in electromagnetism where such effects could be important.

\section{The Classic Image Problem}
\label{imageproblem}

The classic image problem in electromagnetism involves a grounded 
infinite conducting plane and a point charge.  The force of attraction 
between the point charge and the plane is identified using an equivalent
picture of an image charge equidistant behind the plane of opposite sign.   
The field configuration in the half-space of the point charge is mathematically
equivalent to the picture where an image charge of equal magnitude and
opposite sign resides equidistant behind the plane.  Hence, the field
configuration of the point charge and the plane is mathematically equivalent
to the the field between the point charge and the image charge
in the space of interest.

If released from rest, the charge begins to move toward the plane, which
also moves the image charge closer to the midpoint between the charge
and the image.  Consequently, the force of attraction between the charge
and the image rapidly increases.   The expectation in the textbooks
is that Newton's 2nd law can be employed
to write a second-order differential equation to be solved for
$\vec{\bf\/r\/}$({\it\/t\/}).  Typically, one is asked to show that
the time to reach the plane is\cite{griffiths}
\begin{eqnarray}
\tau & = & {\pi{\it\/d\/}\over\/q\/}\sqrt{{\rm\/2}\pi
\epsilon_{0}{\it\/m\/d\/}\ },
\label{tauclassical}
\end{eqnarray}
where {\it\/q\/} is the charge, {\it\/m\/} is the mass and 
{\it\/d\/} is the initial separation between the point charge and
the plane.  The permittivity of free space appears as 
$\epsilon_{0}$ which implies a particular system of units (SI).

\begin{figure}[!h]
\begin{center}
\includegraphics[scale=0.45]{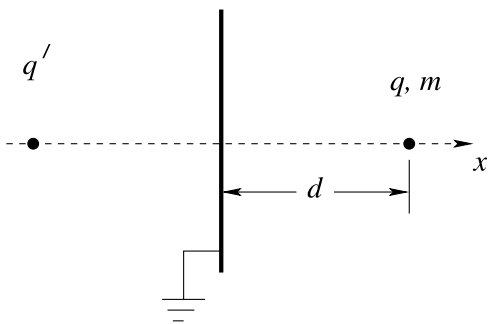}
\caption{Problem setup: Point charge released from rest a distance
{\it\/d} from a grounded infinite conducting plane.}
\label{platepicture}
\end{center}
\end{figure}

Start with a point charge {\it\/q\/} of mass {\it\/m\/} released
from rest a distance {\it\/d\/} from an infinite grounded conducting
plane (see Fig.\ref{platepicture}).  An image charge {\it\/q\/}$^{\,\prime}$
= --{\/\it\/q\/} placed a distance {\it\/d\/} behind the plane will
result in the correct boundary condition for the electric potential
{\it\/V\/}(0) = 0 and the correct field configuration for {\it\/x\/}
$\ge$ 0.  Consequently, by uniqueness theorem arguments, one knows
that the image configuration is mathematically equivalent to the original
problem.  As soon as the original charge moves, so too will the
image charge (very much like two gravitating masses approaching one 
another; however, the image charge is not a physical charge so it
does not introduce kinetic energy into the picture).

The one-dimensional equation of motion under these conditions is
(positive {\it\/x\/} to the right)
\begin{eqnarray}
{-q^{2}\over\/16\pi\epsilon_{0}\/x^{2}} & = & m{d^{\,2}\/x\over\/d\/t^{\,2}}\,,
\label{cleom}
\end{eqnarray}
where the initial conditions in time are {\it\/x\/}(0) = {\it\/d\/} and 
{\it\/v\/}(0) = 0, with the endpoint condition  
{\it\/x\/}($\tau$) = 0. 
Upon integrating once and applying
the initial condition(s) one can write
\begin{eqnarray}
-\,\sqrt{{q^{2}\over\/8\pi\epsilon_{0}\/m\/}\left({1\over\/x\/}-
{1\over\/d\/}\right)} & = & v \ = \ {{d\/x\/}\over\/d\/t\/}\,.
\label{vclassical}
\end{eqnarray}
Already here there is a problem!  When the charge reaches the plane the
value of {\it\/x} becomes 0.  Therefore, the particle's speed becomes
unbounded---clearly unphysical.

Still, if one proceeds to integrate once again to find
an expression relating time {\it\/t\/} and position
{\it\/x\/}, one finds
\begin{eqnarray}
t & = & \tau\left\lbrack
1 - {2\over\/\pi}\,{\rm\/sin}^{-1}\left({\sqrt{x\over\/d}}\,\,
\right)
+ {2\over\pi}\sqrt{{x\over\/d}}\,\sqrt{1-{x\over\/d}}\,\,\right\rbrack\,.
\label{tclassical}
\end{eqnarray}
Setting {\it\/x\/} = 0 identifies the time to 
reach the plane, but one must remember that particle's speed has become
infinite at that moment.

How can this be corrected?  The first step toward a complete
solution is to use the relativistic
expression for the particle's momentum {\it\/p} = 
{\it\/m\/}$\gamma${\it\/v\/}.  Then, changing the time derivative
to a spatial derivative through ${dt = v\/dx\/}$, we can write
\begin{eqnarray}
{-q^{2}\over\/16\pi\epsilon_{0}\/x^{2}} & = & m{d\/(\gamma\/v)\over\/d\/t},
\nonumber\\
{-q^{2}\over\/16\pi\epsilon_{0}\/m}{dx\over\/x^{2}} & = & 
v\,\gamma^{3}\,dv\,.
\label{releom}
\end{eqnarray}

\begin{figure}[!h]
\begin{center}
\includegraphics[scale=0.45]{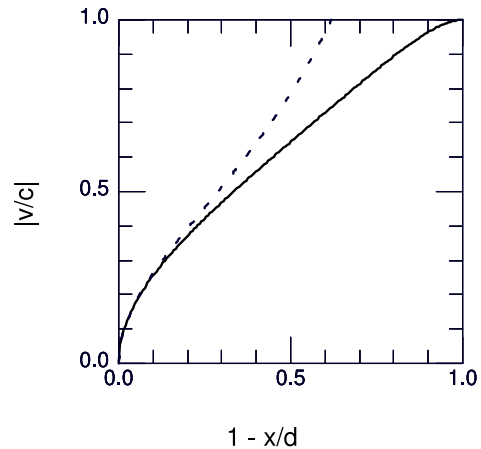}
\caption{Magnitude of velocity divided by {\it\/c\/} versus distance
traveled (that is, 1 minus the position
divided by {\it\/d\/}) for a point charge released 
from rest near a grounded conducting plane. The solid curve 
(Eq.~\ref{relspeed})
describes relativistic dynamics and the dashed curve 
corresponds to nonrelativistic dynamics (Eq.~\ref{vclassical}).}
\label{vversusx}
\end{center}
\end{figure}

\null
The Lorentz factor $\gamma$ = $(1 - v^{2}/c^{\,2})^{-1/2}$ appears
characteristically to modify the dynamics appropriately when the
kinetic energy is not small compared with the rest energy.
This expression can be integrated to find
\begin{eqnarray}
v & = & 
-c\,{\sqrt{\ell^{2}\left(1-{x\over\/d}\right)^{2}+2\ell\/x
\left(1-{x\over\/d}\right)}\over
\ell\left(1-{x\over\/d}\right)+x
}\,, 
\label{relspeed}
\end{eqnarray}
where a characteristic length $\ell$ = 
{\it\/q}$^{\,2}$/(16$\pi\epsilon_{0}mc^{\,2}$) appears.  Notice that 
{\it\/v\/}({\it\/d\,}) = 0
and {\it\/v\/}(0) = $-${\it\/c\/} (with the 
minus sign pointing to the
direction) as it must to avoid superluminal transport. 
For a specific choice of charge, mass, and initial
separation [{\it\/d\/}/({\it\/c}$\tau$) = 0.5, where 
{\it\/d\/}/({\it\/c}$\tau$) = (1/$\pi$)$\sqrt{8\ell/d\/}$],
a comparison is presented in Fig.~\ref{vversusx} of the relativistic
versus nonrelativistic expressions for the speed.
For this particular configuration, the difference seems 
significant.

The next task is clearly to integrate Eq.~(\ref{relspeed}) to
find an expression for {\it\/t\/} as a function of {\it\/x\/}.
Specifically, we must integrate
\begin{eqnarray}
c\/t
& = & 
\int_{x}^{d}
{
\ell\left(1-{x^{\,\prime}\over\/d}\right) + x^{\,\prime}
\over
\sqrt{\ell^{2}\left(1-{x^{\prime}\over\/d}\right)^{2}+2\ell\/x^{\prime}
\left(1-{x^{\prime}\over\/d}\right)}
}\,dx^{\prime}\,. 
\label{tvsxintegral}
\end{eqnarray}
The integral can be done in closed form and written as

\begin{figure}[!h]
\begin{center}
\includegraphics[scale=0.45]{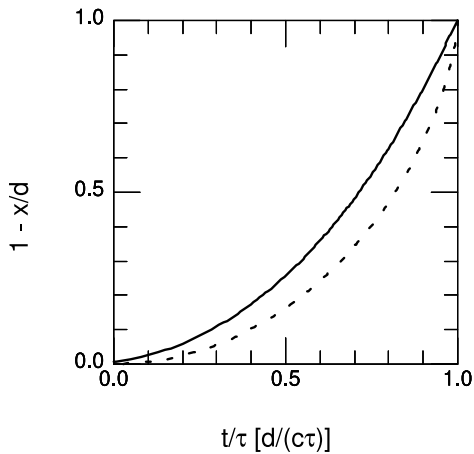}
\caption{Position (distance traveled in units
of {\it\/d\/} is ``1$-${\it\/x/d\/}'') versus time 
for a point charge released from rest
near a grounded conducting plane.   The dimensionless parameter
is chosen to be {\it\/d\/}/({\it\/c}$\tau$) = 0.5. The solid curve corresponds to
relativistic dynamics and dashed curve uses 
the nonrelativistic expression.}
\label{xvst}
\end{center}
\end{figure}

\null
\begin{widetext}
\begin{eqnarray}
{t\over\/\tau} & = & 
\left({d\over\/c\tau}\right)^{2}\left({\pi\over\/4\/}\right)
\left\lbrace
\left\lbrack
{1\over\sqrt{2}}\left(\sqrt{\ell\over\/d}\right)
\left({d\over\/\ell}-1\right)
\right\rbrack
\right.
\nonumber\\
& \ & 
\left.
\cdot
\left(1 -
\sqrt{{1\over\/8}\left({\ell\over\/d}\right)\left({2d\over\/\ell}
-1\right)^{2}+1-\left(
{\sqrt{2}\/x\over\/d\/}\sqrt{d\over\ell}-{\sqrt{2}\over\/4}\left(
{\ell\over\/d}\right)\left({2\/d\over\ell}-1\right)
\right)^{2}}\,
\right)
\right.
\nonumber\\
& \, &
\left.
+\left\lbrack
{\ell\over\/4d\/}\left({2d\over\ell}-1\right)\left({d\over\ell}-1\right)+1
  \right\rbrack
\right.
\nonumber\\
& \ & 
\left.
\left\lbrack  
{\rm\/sin\/}^{-1}\left({
{x\over\/d\/}\sqrt{d\over\ell}-{1\over\/4}\sqrt{\ell\over\/d}\left(
{2d\over\ell}-1\right)
\over
\sqrt{{1\over\/16}
\left({\ell\over\/d}\right)\left({2d\over\ell}-1\right)^{2}+{1\over\/2}}
}\right)
-
{\rm\/sin\/}^{-1}\left( 
{
-{1\over\/4}\left({2d\over\ell}-1\right)
\over
\sqrt{{1\over\/16}
\left({2d\over\ell}-1\right)^{2}+{1\over\/2}{d\over\ell}}
}
\right)
\right\rbrack
\right\rbrace\,.
\quad\quad\quad\quad
\label{trelativistic}
\end{eqnarray}
\end{widetext}
A position-time analysis is now possible by comparing Eqs.~(\ref{tclassical})
and (\ref{trelativistic}).  The comparison is shown in Fig.~(\ref{xvst})
with the solid curve representing relativistic dynamics and the dashed
curve displaying the nonrelativistic expression.  The same choice is made
here for the dimensionless parameter as was made
previously [namely, {\it\/d\/}/(c$\tau$) = 0.5].

The slope of the tangent line is of course a measure of the magnitude
of the instantaneous velocity.  However, it cannot be read off directly 
since there is a
factor of {\it\/d\/}/($\tau${\it\/c}) in the ratio.  And yet, notice that the 
dashed curve exhibits a slope for its
tangent line near {\it x\/} = 0 that is very large.  Numerically
the magnitude of the velocity 
exceeds 1 (in units of {\it\/c\/}) for {\it\/t\/}/$\tau$ 
$\ge$ 0.62
in the dashed curve.
At {\it\/t\/}/$\tau$ =  0.8 the speed exceeds 2, and it exceeds 4 at
{\it\/t\/}/$\tau$ of 0.96.   Furthermore, we point out that the 
relativistically
consistent curve displays a slope corresponding to a speed less than 1 
throughout, only does the speed approach 1 at the point 
{\it\/x\/} = 0.

Now that we have the expression for {\it\/t\/} as a function of
{\it\/x\/}, it is a straightforward exercise to take {\it\/x} 
to zero and identify the time to reach the plane. 
By setting {\it\/x\/} $\rightarrow$ 0, then $t\rightarrow\/T$, and  
we have,
\begin{eqnarray}
{T\over\tau} & = & 
\left({d\over\tau\/c}\right)
\left\lbrace
{1\over\/2}
-\left({d\over\tau\/c}\right)^{2}
{\pi^{2}\over\/16\/}
\right.
\nonumber\\
& \ &
\left.
+\left\lbrack
{1\over\pi}\left({\tau\/c\over\/d}\right)
+\left({d\over\tau\/c}\right){\pi\over\/16}
+{\pi^{3}\over\/128}\left({d\over\tau\/c}\right)^{3}
\right\rbrack
\right.
\nonumber\\
& \ &
\left.
\cdot
\left\lbrack
{\pi\over\/2}
-
{\rm\/sin\/}^{-1}\left({{\pi^{2}\over\/8}
\left({d\over\tau\/c}\right)^{2}-2\over
{\pi^{2}\over\/8}
\left({d\over\tau\/c}\right)^{2}+2
}\right)
\right\rbrack
\right\rbrace\,,\,
\end{eqnarray}
where $\tau$ is the original nonrelativistic
expression given by Eq.~(\ref{tauclassical}).  We remark that
if $c\rightarrow\infty$, the right hand side approaches 1, so
{\it\/T\/}$\rightarrow\tau$.

A comparison between the nonrelativistic expression for the time to reach
the plane as compared with the fully relativistic expression is
ultimately useful to be able to identify the conditions under which
the original textbook expression fails.  This comparison is deferred
until after a discussion of the field-retardation effects.

\section{Field Retardation Effects}
\label{lienardwiechart}

The discussion of the previous section improves upon the nonrelativistic
analysis which is typically used to find the time to reach the plane.  
However, it too is lacking in some respects.  When the charge is a distance
{\it\/x} from the plane and moving toward the plane, the force
it experiences is different from a static Coulomb picture because the
``electromagnetic news" travels at the speed of light.  Hence, the
electric field it experiences must be related to the electric and
magnetic potentials at some earlier time (the so-called retarded
time).  Simply put, it takes a time {\it\/x\//c} for information
on the particle dynamics to reach the plane.  The charge
configuration on the plane is updated (which means the image charge location is
updated) and then that information requires a time {\it\/x\//c} to
get back to the point charge.  This type of physical circumstance
is describable by the Li\'enard-Wiechart 
potentials for a moving point charge\cite{jackson,griffiths2}.

Suppose the charge is located at position {\it\/x\/} and moving
with a velocity $-${\it\/v\/} (again, positive {\it\/x\/} to the right).  
The general problem to calculate
the retarded potentials is nontrivial, but when the motion
is one-dimensional it simplifies significantly.  In particular, the
electric field experienced by the charge in this case is reduced by
two powers of $\gamma$.  It can therefore be written as
\begin{eqnarray}
E & =& {-q\over\/4\pi\epsilon_{0}}\left({1\over\/4x^{2}}\right)
\left(1-v^{2}/c^{2}\right).
\label{retardedfield}
\end{eqnarray}
The force drawing the charge toward the plane is therefore velocity
dependent and it is reduced as compared with the static case.  Consequently,
the relativistic and properly time-delayed equation of motion is
\begin{eqnarray}
{-q^{2}\over\/16\pi\epsilon_{0}\/x^{2}}
\left(1-{v^{2}\over\/c^{\,2}}\right) & = & m{d(\gamma\/v)\over\/d\/t}\,.
\label{retardedeom}
\end{eqnarray}
Remarkably, this too can be integrated to find the velocity as a function
of position.  It becomes
\begin{eqnarray}
{dx\over\/dt} \, = \, v & =& 
-\,c\,\sqrt{1-\left({xd\over\/xd+3\ell(d-x)}\right)^{2/3}}\,.
\label{retardedvel}
\end{eqnarray}
Another integral is required to identify a relationship between 
position and time in this relativistically consistent and properly 
casual description of the dynamics.  Unfortunately, a closed form
expression has not been found.  Instead, the integral will be carried
out numerically for the case of interest here.  That is, the
time {\it\/T\/} to reach the plane will be computed
and displayed graphically to help clarify the size of
the effects.  In particular, the time {\it\/T\/} is
\begin{eqnarray}
{T\over\tau} & = & 
\,\int_{0}^{1}\,{   
\left({d\over\tau\/c\/}\right)
dz\over
\sqrt{1-\left(
{z\left/ \left\lbrack\/z+{3\pi^{2}
\over\/8}\left({d\over\tau\/c}\right)^{2}(1-z)\right\rbrack\right.}
\right)^{2/3}}}\,.\quad
\label{finaltime}
\end{eqnarray}
Time {\it\/T\/} is normalized here by the original nonrelativistic time.  
A ratio
of 1 indicates that the original nonrelativistic expression is completely 
satisfactory.
Deviations from 1 indicate that relativistic effects and/or field
retardation effects become important.

\begin{figure}[!h]
\begin{center}
\includegraphics[scale=0.45]{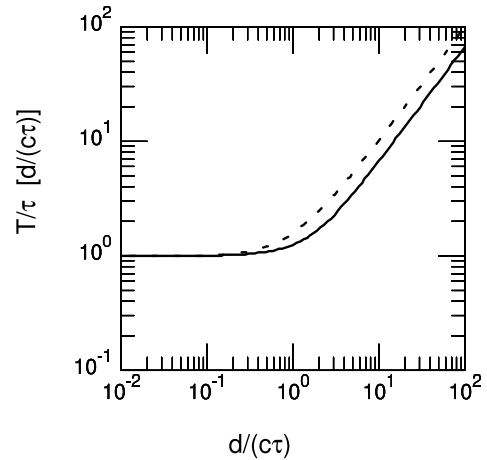}
\caption{Time to reach the plane versus scaled initial separation.  
Solid curve includes relativistic corrections, dashed curve includes 
relativistic correction plus retarded (Li\'enard-Wiechart) potentials.}
\label{tversusdotc}
\end{center}
\end{figure}

The time to reach the plane is again a function of the dimensionless
parameter {\it\/d\/}/($\tau${\it\/c\/}).  In Fig.~(\ref{tversusdotc})
the full solution is plotted as a function of 
{\it\/d\/}/($\tau${\it\/c\/}).  One must keep in mind that the dimensionless
parameter is related to the charge, mass, and initial separation through
\begin{eqnarray}
{d\over\tau\/c\/} & = & {1\over\pi}\sqrt{8\ell\over\/d}
\quad = \quad {q\over\pi\sqrt{2\pi\epsilon_{0}mc^{2}d}}\,.
\label{dotc}
\end{eqnarray}

Graphical results indicate that so long as the dimensionless
parameter {\it\/d\/}/($\tau${\it\/c\/}) is less than roughly 0.1, then
the nonrelativistic expression for the time is completely adequate.  For
values of the parameter greater than 1 the relativistic and
retardation effects become very important.  What remains to be
considered is to identify the physical situations that correspond to 
each of these regimes.

\section{Discussion}
\label{discuss}

There is of course some concern about using a solution for 
$\tau$ that involves
superluminal transport.  However, we have identified the
crossover point where the relativistic expression must be
used (see Fig~\ref{tversusdotc}).  If a particular application
falls to the left of the crossover point on Fig.~(\ref{tversusdotc}), 
then the nonrelativistic
formulation is completely adequate.  It will be useful
to consider a few example configurations to get a feel for 
this dimensionless parameter {\it\/d\/}/($\tau${\it\/c\/}).
Typical laboratory-type charges are microcoulombs, with typical masses
in the milligram-to-gram range, and initial separations in the
neighborhood of tens of centimeters.  

For the typical configuration described above, the
important parameter comes out to be of the order 10$^{-6}$
as can be seen from the information in Table~\ref{dptable}.  Consequently,
the nonrelativistic solution is completely sufficient for this case.
We emphasize that even though the nonrelativistic expression for the
total time to reach the plane is valid, it does not mean the particle
was completely nonrelativistic.
Rather, the time spent moving relativistically was such a small
part of the total time that relativistic corrections are unimportant.
It is especially reassuring to notice that the number 10$^{-6}$ is several
orders of magnitude smaller than the region in Fig.~(\ref{tversusdotc})
where the relativistic and retardation effects begin to be important.

\begin{table}[!h] 
\begin{tabular}{|l|l|l|l|}
\hline
$q$ & $d$ & $m$ & ${d\,/(\tau\/c})$ \\
\hline
 1$\mu$C & 1 (cm) & 1 milligram & 
$\sim$ 10$^{-6}$\\
electron & 0.1 (meter) & $m_{e}$ & 
7.5$\times$10$^{-8}$\\
electron & 1 (fm) & $m_{e}$ & 
$\sim$ 1 \\
proton & 10 (fm) & $m_{p}$ & $\sim$ 6$\times$10$^{-3}$\\
\hline
\end{tabular}
\caption{Comparison of different physical configurations and
the resulting dimensionless parameter $d/(\tau\/c)$ which determines
the region of applicability of relativity and field-retardation effects.}
\label{dptable}
\end{table}

Next we explore the case of an electron released from rest a
distance 10 cm from a grounded conducting plane.  Again from the
table we extract the important dimensionless parameter for this
case to be a number of the order of 10$^{-8}$ (even smaller than
the previous case!).  This too is completely legitimately described 
by nonrelativistic formulae.
The argument that saves the nonrelativistic expression is that even though
the last tiny fraction of the trip to the grounded plane is superluminal, 
the charge spends so little time moving faster than light that the
the deviation from the ``correct" time is immeasurably small.  So, once
again, the nonrelativistic expression is sufficient.

Lastly, we investigate extreme cases where relativistic and retardation
effects could be important.  Consider an electron released from
rest a distance 1 femtometer from a grounded conducting plane.
The important parameter determining the dynamics is again the
number {\it\/d}/($\tau${\it\/c}) which comes out to be of the order 1 for this
configuration.  The nonrelativistic expression will fail noticeably
for this case.  Not only is relativity important for the dynamics, but 
field retardation effects are very important.  A proton
released 10 fm from the plane results in a parameter
{\it\/d\/}/($\tau${\it\/c}) of 10$^{-3}$.

Before concluding, it is useful to extend the solution presented
above for the electrical charges to the case of two gravitating point 
masses.  Consider two point masses
{\it\/m\/} initially separated by a distance 2{\it\/d\/}.  We have immediately
solved that problem as well using a correspondence between charge
and mass, and between the inverse of the permittivity of free space and
Newton's universal gravitational constant.
The above dynamical
expressions relating position, velocity and time
hold for the case of gravity while the relevant dimensionless parameter
becomes
\begin{eqnarray}
{d\over\tau\/c\/} & = & {\sqrt{2}\over\pi}\sqrt{{mG\over\/d\/c^{\,2}}}\,,
\label{dparamgravity}
\end{eqnarray}
where {\it\/G\/} is Newton's universal gravitational constant.
For most configurations this parameter is extremely small. Hence, once
again, the nonrelativistic expression for the time to reach the center
of mass is completely safe to use.  

\section{Conclusion}
\label{summary}

The classic image problem of a point charge near a grounded 
conducting plane invites the notion of a mathematical tool of
replacing the grounded plane with a fictitious charge (a so-called image
charge) which, together with the original charge, produces the same
electric potential in the region of interest and respects the boundary
conditions.  The existence and uniqueness theorems indicate that
the scalar potential and therefore the (vector) electric fields must
be equivalent.  It is a beautiful solution to an otherwise difficult
problem.

The follow-up problem involves computing the time it takes the charge
to reach the plane when the charge is released from rest.  A straightforward
(although incorrect) analysis from nonrelativistic classical mechanics 
results in
an expression for the time which is proportional to the initial 
separation to the three-halves power, also roportional to the square root
of the mass, and inversely proportional to the charge.  We have pointed
out in this article the serious formal flaw in that proposed solution.
The flaw is linked to the observation that the velocity of the charged 
particle exceeds the speed of light for a small portion of its trip toward 
the grounded plane.  It is very important to sort out
when the solution is still valid and acceptable, and when it
is noticeably inadequate.  This has been the purpose of our article.

We have shown that the relativistic version of the problem is completely
tractable and results in subluminal dynamics which changes the expression
for the time to reach the plane by monotonically increasing the result
as a function of the dimensionless parameter {\it\/d\/}/($\tau${\it\/c\/}).

A more sophisticated approach to point charge dynamics near a
grounded conducting plane must include field-retardation effects.
This has also been added to the solution discussed in this paper.
Using time-delayed fields (Li\'enard-Wiechart potentials)
increases further the time required for the particle
to reach the plane.  Both effects,
relativistic and retardation influences, are easily understood in
terms of their tendency to increase the time for the
particle to reach the plane.

Finally, we briefly mention other problems where these same effects
could be important.  First, a point charge {\it\/q\/} of mass
{\it\/m\/} near a fixed line charge
$-\lambda$ experiences a force that is proportional to 1/{\it\/r\/}. 
The nonrelativistic time to reach the line charge can be computed 
as\cite{integral1} $\tau$ =
$\pi$\/{\it\/d\/}\/$\sqrt{\epsilon_{0}\/m\//(\lambda\/q)}$.
However, the charge goes superluminal as it reaches the line charge.
A point dipole {\it\/p} of mass {\it\/m\/} near a fixed line charge $-\lambda$
experiences a 1/{\it\/r\/}$^{\/2}$ force.  Using nonrelativistic dynamics
again we find the time to reach the line charge is $\tau$ = 
($\pi$\/{\it\/d}\//4)$\sqrt{\pi\epsilon_{0}\/m\/d\//(\/p
\lambda)}$.
Furthermore, 
a point charge {\it\/q\/} of mass {\it\/m\/} near a point 
dipole {\it\/p\/} of mass {\it\/m\/} will experience an inverse cube 
force ($1/r^{3}$).  
The nonrelativistic time for the two to meet at the midpoint (each of which
is initially a distance {\it\/d\/} away from the midpoint) is $\tau$ = 
${\it\/d\/}^{\,2}\sqrt{2\,\pi\epsilon_{0}\/m/(\/p\/q\,)}$.
And lastly, two identical point dipoles {\it\/p\/} with mass {\it\/m\/}
will experience a force proportional to 1/{\it\/r\/}$^{\,4}$.  
The nonrelativistic
expression for the time to reach the midpoint is\cite{integral4} $\tau$ =
(2/3)$\kappa${\it\/d}$^{\,2}$\,
$\sqrt{\pi\epsilon_{0}\/m\/d\,\,}$/$p$,
where the constant $\kappa$ can be computed as
\begin{eqnarray}
\kappa & = & \int_{0}^{\pi/2}\,\left[{\rm\/sin\/}^{2/3}
\theta\,\right]\/d\theta
\quad = \quad 1.12\ldots
\label{sineintegral}
\end{eqnarray}
In each case above, analytic or numerical
expressions for velocity as a function of position and position as 
a function of time can be identified.  Relativistic corrections and
field retardation effects are equally readily obtainable.  For increasingly
steeper {\it\/r\/}-dependence, one might expect that the effects
could be increasingly more important.  The small-distance 
behavior determines the relative importance of the effects, and hence 
the dipole-dipole is the most sensitive to these effects.

\begin{acknowledgments}
This work was supported in part by the National 
Science Foundation under grant No. PHY-0555521.
\end{acknowledgments}

\end{document}